\documentclass[doublecolumn,10pt]{IEEEtran}
\usepackage[columnwise]{lineno}
\usepackage{amssymb,amsmath}
\usepackage[colorlinks,linkcolor=red,citecolor=cyan,backref=page, linktocpage]{hyperref}
\usepackage{graphicx} 
\usepackage{booktabs}  
\usepackage{multirow}  
\usepackage{amsmath}   
\usepackage{xcolor}    
\title{Spectral point transformer for significant wave height estimation from sea clutter}
\author{Yi Zhou, Li Wang, Hang Su, Tian Wang
\thanks{
Yi Zhou is  with the Department of Electronic Information Engineering, Dalian Maritime University, Dalian, 116026, China. Email: (yi.zhou@dlmu.edu.cn).
Li Wang is with the Department of Software Development, SVA Communication Technology Co., Ltd., 200233, Shanghai, China. Email: (wang\_li\_apple@163.com).
Hang Su is with the Department of Computer Science and Technology, Tsinghua University, Beijing, 100084, China. Email:(suhangss@mail.tsinghua.edu.cn). 
Tian Wang is with the School of Artificial Intelligence, Beihang University, Beijing, 100083, China. Email:(wangtian@buaa.edu.cn).
}
}
\graphicspath{{../figure/}, {./figure/}}
\begin{document}
\maketitle
\begin{abstract}
This paper presents a method for estimating significant wave height ($H_s$) from sparse \underline{S}pectral \underline{P}oint using a \underline{T}ransformer-based approach (SPT). Based on empirical observations that only a minority of spectral points with strong power contribute to wave energy, the proposed SPT effectively integrates geometric and spectral characteristics of ocean surface waves to estimate $H_s$  through multi-dimensional feature representation. The experiment reveals an intriguing phenomenon: the learned features of SPT align well with physical dispersion relations, where the contribution-score map of selected points is concentrated along dispersion curves. Compared to conventional vision networks that process image sequences and full spectra, SPT demonstrates superior performance in $H_s$ regression while consuming significantly fewer computational resources. On a consumer-grade GPU, SPT completes the training of regression model for 1080 sea clutter image sequences within 4 minutes, showcasing its potential to reduce deployment costs for radar wave-measuring systems. The open-source implementation of SPT will be available at \url{https://github.com/joeyee/spt}
\end{abstract}

\section{Introduction}
Significant wave height ($H_s$) is a critical statistical parameter for describing sea state, and its accurate estimation remains a focal point in physical oceanography and remote sensing. Traditional wave measurement techniques, such as wave buoys and acoustic Doppler current profilers (ADCPs), although highly precise, are limited to providing time series data at single point. These methods fail to capture the spatial distribution of waves and require high maintenance costs, making them impractical for extensive deployment over large areas.

In contrast, X-band marine radar, as an active microwave remote sensing device, offers advantages such as all-weather operation, spatial resolution on the order of meters, temporal resolution at the second level, and a wide coverage range up to several kilometers. Although originally designed for ship navigation and collision avoidance, the sea clutter signals received by radar contain rich information about ocean surface dynamics. Since the 1980s, inverting wave parameters (e.g., wave height, period, direction, and current speed) from X-band radar image sequences has become an important branch of marine remote sensing, giving rise to commercial systems like WaMoS II \cite{HessnerBook2008}.

However, it is challenging to derive sea wave heights from the gray intensity of radar images . This is not only a nonlinear and ill-posed mathematical inversion problem, but also influenced by complex factors such as radar imaging mechanisms (tilt modulation, shadowing effects, hydrodynamic modulation), environmental conditions (wind speed, rain interference) and hardware parameters (antenna height, magnetron aging). In recent years, advances in computational power and the integration of artificial intelligence have transformed research paradigms in this field. The transition has been from early Signal-to-Noise Ratio (SNR) \cite{borge2004,borge2008} to uncalibrated geometric methods \cite{Gangeskar2014,Navarro2022}, and now to data-driven deep learning approaches \cite{swhformer2024,3dvgg2023}.

Both traditional methods (SNR and geometry) rely on  expert's feature engineering. However, radar images contain richer information about the  evolution of textures and shapes that these methods are unable to fully exploit. With the rise of deep learning, end-to-end inversion approaches based on general vision models -- convolutional neural networks (CNNs) \cite{vgg2015} and vision transformers \cite{vit2021} -- have emerged as alternatives. These methods leverage transfer learning by pretraining models on large non-marine visual datasets and fine-tuning them on small marine clutter image or sequence datasets. Despite their advantages, these approaches face several drawbacks: the extensive pre-training data primarily consist of non-oceanic visual images, which diverts network parameters away from ocean-specific features. This reduces sensitivity to wave-related visual cues and increases the risk of overfitting during fine-tuning on limited marine data. Additionally, the high computational demands and long pre-training cycles for general visual models make them less adaptable for real-time updates in dynamic environments.

Can we leverage prior knowledge about wave energy in sea clutter to design an efficient network for significant wave height regression? In \cite{senet2001, borge2008}, threshold values in sea clutter spectra were used to distinguish between wave energy and background noise, revealing a maximum difference of two orders of magnitude between the two. Motivated by this observation, we propose collecting the top $N$ spectral points located within one percent of the power maximum as representative of the wave energy.

\begin{figure}[!ht]
  \centering
  \includegraphics[width=1.\linewidth]{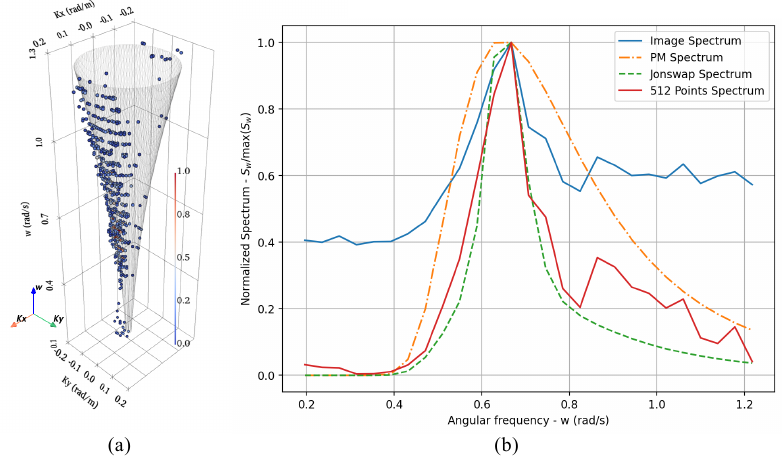}
  \caption{(a) Top 512 sea clutter spectral points in the wavenumber-frequency domain and the underlying dispersion shell. (b) Energy distribution of the sparse points compared with the empirical PM and JONSWAP wave spectra.}
  \label{fig.spectrum}
\end{figure}

Figure \ref{fig.spectrum} illustrates that for a sea clutter sequence with $H_s = 6.3\,\text{m}$ and peak angular frequency $\omega_p = 0.65\,\text{rad/s}$, most of the top 512 spectral points lie near the surface of the dispersion shell. 
In the maximum-normalization spectrum plot (Fig. \ref{fig.spectrum}  b), the  image spectrum denotes the amplitude of the 3D Fast Fourier Transformation (3D-FFT) applied to the image sequence of sea clutter.
Top 512 spectral points are extracted from the image spectrum and then integrated into a spectrum by accumulating their power in every frequency bin.
The energy distribution of these points aligns well between empirical Pierson–Moskowitz (PM) \cite{pierson1964} and JONSWAP spectrum \cite{jonswap1973}, suggesting a strong intrinsic relationship between sparse high-energy spectral points and wave spectrum.

Based on this insight, we introduce a Spectral Point Transformer (SPT) network to regress significant wave height using the energy magnitude and spatial distribution (wavenumber-frequency coordinates) of high-value spectral points. By leveraging prior knowledge of wave spectrum characteristics, SPT significantly reduces input data volume compared to conventional CNNs \cite{vgg2015} or vision transformers (ViT) \cite{vit2021}, enhancing training efficiency.

The contributions of this work are summarized as follows:
\begin{enumerate}
    \item SPT employs sparse spectral points of sea clutter to regress $H_s$ through deep learning, reducing reliance on empirical modulations and enhancing regression accuracy. To our knowledge, this work presents the first neural network trained to retrieve significant wave height directly from sparse points in spectrum. Notably, the experiment reveals that the features learned by SPT exhibit the consistency with physical dispersion relations.
    
    \item Experimental results demonstrate that SPT achieves higher regression precision compared to state-of-the-art visual regression models while requiring significantly less computational resources. For instance, processing 1080 sea clutter sequences (64 frames of $128 \times 128$ resolution images) consumes only 1098 Mb graphical memory and can be trained in 4 minutes on a single RTX5090 GPU for 300 epochs. This lightweight and efficient regression network has great potential to reduce costs in deploying wave inversion systems at the front-end radar equipment.
\end{enumerate}

\section{Related Works}

X-band marine radar emits electromagnetic waves (approximately 3 cm in wavelength), which interact with the sea surface primarily through Bragg resonance. The bright and dark stripes observed in radar images do not directly correspond to wave peaks and troughs but are a result of long gravity waves modulating shorter ripples \cite{ward2006}.
Nieto Borge and colleagues established a standard three-dimensional spectral analysis procedure to separate sea wave signals from noise in the wavenumber-frequency domain, forming the cornerstone of this field \cite{young1985,borge2004,borge2008}. While this method has been widely adopted in commercial systems like WaMoS II  \cite{HessnerBook2008}, its strong dependency on external calibration data and failure under specific environmental conditions remain critical limitations. In contrast, our proposed SPT method leverages the high-energy spectral point information from sea clutter images. By constructing a transformer network based on point cloud regression, we establish a data-driven mapping relationship between sparse spectral points and $H_s$. This approach reduces reliance on empirical numerical formulas while improving the accuracy of significant wave height estimation.

Conventional deep learning-based methods typically introduce advanced visual classification models for significant wave height estimation from sea clutter images.  In the literature \cite{DuanRS2020, 3dvgg2023, zuo2024, rezvov2024, swhformer2024}, along with the increasing of the computing power, VGG\cite{vgg2015}, ResNet\cite{resnet2016}, ViT\cite{vit2021} are utilized to estimate the significant wave heights respectively or jointly. However, directly applying these general visual models to sea clutter images has several drawbacks: it wastes computational resources and may introduce errors caused by training on non-oceanic datasets. Additionally, the large number of model parameters limits the ability to update weights flexibly at radar endpoints. In contrast, our SPT method is based on the prior knowledge that sparse spectral points with strong intensities are intrinsically related to the wave's energy indicator $H_s$. By constructing a transformer network for regression using this principle, we effectively reduce redundant information in image sequences and lighten the computational burden. As a result, even with a lightweight transformer architecture, our method achieves better performance than general-purpose visual networks.

PointNet \cite{pointnet2017} and PointNet++\cite{pointnet2} established the foundational paradigms for global and hierarchical feature extraction from point cloud. The inclusion of transformer in PCT (point cloud transformer) \cite{pct2021} reflects the recent shift toward attention mechanisms for superior global-local context modeling. The PCT method employs geometric features such as normal parameters and local curvature, along with 3D coordinates, as input features. Although these points are discretely distributed, they exhibit continuous geometric correlations. Therefore, in designing embedding modules for PCT, convolution are often used to associate neighboring points and enhance feature dimensions. Given that the coordinate values and feature magnitudes of samples are consistent across the same numerical range, batch normalization (BatchNorm) is typically applied during normalization. For networks based on spectral points, we consider both the spectral energy $p$ and the spatial position of each point as inputs. The key challenge lies in handling the spatial independence between spectral points and the significant differences in their energy levels. In designing our SPT network for sparse high-energy spectral points, we adopt a multi-layer perceptron (MLP) embedding and meticulous normalization to enhance regression performance. 

\section{Methodology}
\begin{figure}[!ht]
    \centering
    \includegraphics[width=1.\linewidth]{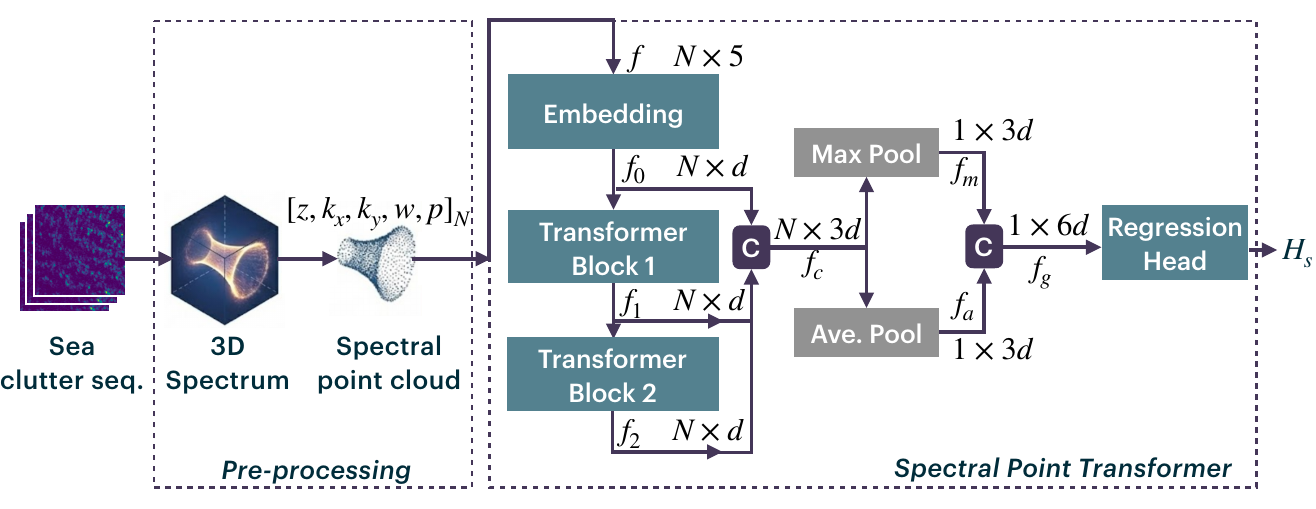}
    \caption{The framework of Spectral Point Transformer (SPT).}
    \label{fig.SPT}
\end{figure}

In this section, we present the methodology of the proposed approach, focusing on the pre-processing steps and the SPT network architecture. Figure \ref{fig.SPT} provides an overview of the system framework.

\subsection{Pre-processing}
The input consists of sea clutter image sequences captured under varying sea states. To extract the most relevant features for effective wave height estimation, we perform the following pre-processing steps to obtain the spectral point cloud.

First, raw sea clutter image sequences are processed through a 3D-FFT. Second, in the resulting power spectrum, the highest $N$ spectral points with the strongest amplitude are selected. Third, the corresponding wave number-frequency coordinates $(k_x, k_y, \omega)$, amplitude values $p$, and clutter's zone index $z$, which takes values in $\{1,2,3\}$, are used as initial features for point cloud representation: $f = [z, k_x, k_y, \omega, p]_N$. The dimension of $f$ is $N \times 5$.

This design takes advantage that the spectral energy is concentrated on the dispersion surfaces, which consist of discrete spectral points due to resolution limitations (as shown in Figure \ref{fig.spectrum}). Based on the observation that wave-induced energy exceeds background noise by approximately two orders of magnitude~\cite{senet2001}, we collect data from the top $N$ spectral points located between the power maximum and one percent of its value to represent the wave energy. After pre-processing, only about 0.05\% of the original image sequence spectrum data is retained, achieving a compression ratio of approximately 2048 times. This significantly reduces computational requirements for subsequent model training.

\subsection{SPT Network Architecture}
The SPT network consists of three main components: an embedding module, stacked transformer blocks, and a regression head. Each component addresses specific challenges posed by the spectral point cloud under varying conditions.

\subsubsection{Embedding Module}
In the input feature $f$, clutter's zone index $z$, wavenumber coordinates $(k_x, k_y)$, and angular frequency $\omega$ do not vary at equal geometric intervals. Traditional sine/cosine positional embeddings used in standard transformers~\cite{attention2017} are unsuitable for handling such unequal spacing. 
Therefore, both the positional information and the amplitude of a spectral point are equally important as initial token features. By omitting explicit positional embedding, the SPT is encouraged to autonomously learn and emphasize discriminative spectral coordinates from the hidden feature representations.

The selected spectral points, being intensity-sorted, are discontinuous in spectral space. Using 1D-convolution embedding (e.g., \cite{pct2021}) would integrate wrong neighborhoods that attenuate isolated strong points via false adjacency. We therefore adopt a two-layer MLP to embed token features into higher dimensions instead. The first layer of the MLP is expressed as follows:
 \begin{equation}
    \textrm{Layer}(f)=\textrm{SiLU}\left(\textrm{LayerNorm}\left( \textrm{Linear}(f)\right)\right).
\end{equation}
Here, SiLU denotes Sigmoid-weighted linear units (SiLU)\cite{silu}, LayerNorm calculates statistics across the feature dimension for each individual sample independently as \cite{attention2017} and Linear denotes the matrix multiplication to change the feature dimension. Let $f, f_0$ denote the input and output feature of the embedding module, the whole expression is written as:
\begin{equation}
    f_0 = \textrm{Layer}_2\left(\textrm{Layer}_1(f)\right).
\end{equation}

Here, $f_0\in R^{N \times d}$, $d$ means the embedding dimension. The inner $\textrm{Layer}_1$ first increases the dimension $f$  from 5 to $d/2$, then the outer $\textrm{Layer}_2$ further increases the intermediate dimension to $d$.

\subsubsection{Transformer Block}
Each transformer block consists of two sub-layers: the self-attention module and a feed-forward network (FFN) with two-layer MLP. To address the challenge of large dynamic range in power across spectral points for different sea states (where amplitude differences between high and low sea states can reach several orders of magnitude), this paper incorporates Query-Key normalization (QK Norm.) \cite{qknorm2020} into block design to enhance model performance.

The self-attention mechanism is at the core of the transformer architecture. It enables the model to connect other tokens in the sequence by assigning different attention scores weighted by their relevance. Let $\mathbf{X} \in \mathbb{R}^{N \times d}$ denote the input matrix, where $N$ is the sequence length and $d$ is the embedding dimension. It projects the input into query , key, and value spaces.

Define projecting matrices $\mathbf{W}_Q, \mathbf{W}_K, \mathbf{W}_V \in \mathbb{R}^{d \times d}$ as follows:
\begin{equation}
\mathbf{Q} = \mathbf{X}\mathbf{W}_Q, \quad \mathbf{K} = \mathbf{X}\mathbf{W}_K, \quad \mathbf{V} = \mathbf{X}\mathbf{W}_V,
\end{equation}
where $\mathbf{Q}, \mathbf{K}, \mathbf{V} \in \mathbb{R}^{N \times d}$. 
For computational efficiency, the three operations are typically combined into a single matrix multiplication:
\begin{equation}
\text{Proj}(\mathbf{X}) = [\mathbf{Q}, \mathbf{K}, \mathbf{V}] = \mathbf{X} \cdot [\mathbf{W}_Q, \mathbf{W}_K, \mathbf{W}_V]_{N \times 3d}.
\end{equation}

Given the query, key and value matrix, the standard scaled dot-product attention is defined as:
\begin{equation}
\text{Sa}(\mathbf{Q}, \mathbf{K}, \mathbf{V}) = \text{softmax}\left( \frac{\mathbf{Q}\mathbf{K}^T}{\sqrt{d}} \right) \mathbf{V}.
\label{eq.sa}
\end{equation}

When the dot product of $\mathbf{Q}$ and $\mathbf{K}$ is excessively large, the softmax operation produces a highly peaked probability distribution (approaching one-hot), leading to the gradient vanishing and unstable training. To mitigate the impact of high-energy features on gradient stability during backpropagation, we adopt QK Norm. scheme \cite{qknorm2020} . Specifically, before computing the inner product of $\mathbf{Q}$ and $\mathbf{K}$, we perform the Root Mean Square Normalization (RMSNorm) \cite{rmsnorm2019} on each row vector $\mathbf{q}_i \in \mathbb{R}^d$ of $\mathbf{Q}$:
\begin{equation}
\hat{\mathbf{q}}_i = \frac{\mathbf{q}_i}{\sqrt{\frac{1}{d} \sum_{j=1}^d q_{ij}^2 }}.
\end{equation}

After normalizing all $N$ rows, we obtain $\hat{\mathbf{Q}}$. Similarly, $\mathbf{\hat{K}}$ and $\mathbf{\hat{V}}$ are computed. This is denoted as:
\begin{equation}
[\hat{\mathbf{Q}}, \hat{\mathbf{K}}, \hat{\mathbf{V}}] = \text{RMSNorm}([\mathbf{Q}, \mathbf{K},  \mathbf{Q}]).
\end{equation}

Substituting into \eqref{eq.sa} yields the  output of the attention module as:
\begin{equation}
\text{Attn}(\mathbf{X}) = \text{Sa}\left( \text{RMSNorm}\left(\text{Proj}\left(\mathbf{X} \right)\right)\right).
\end{equation}

Notably, transformer models employ multi-head attention, dividing features into multiple subspaces for parallel attention operations within each subspace. The outputs from all heads are concatenated to enhance  the learning of feature diversity. 

Finally, pre-norm (Layer normalization) is applied to both attention and MLP inputs, with residual connections \cite{residual2016}. Here, the two-layer MLP first increases the token's feature dimension to $4d$ and then projects back to the original $d$  in the conventional way as \cite{attention2017}.
For a given input feature $f_{i-1} \in \mathbb{R}^{N\times d}$ to the $i$-th transformer block, the output feature $f_i$ is computed as:
\begin{equation}
\begin{cases}
f_i' &= f_{i-1} + \textrm{Attn}\left(\textrm{LayerNorm}(f_{i-1})\right), \\
f_{i} &=  f_i' + \textrm{MLP}\left(\textrm{LayerNorm}(f_i')\right).
\end{cases}
\end{equation}

As shown in Figure \ref{fig.SPT}, two transformer blocks are stacked, their outputs being denoted $f_1$ and $f_2$. These are concatenated with the output of the embedding module $f_0$ to form an intermediate variable $f_c \in \mathbb{R}^{N\times 3d}$. The max and average-pooling operations compute the maximum and mean values across $N$ data points for each dimension of the feature , resulting in the features $f_m$ and $f_a$ of size $1\times 3d$. Concatenating them yields the final global feature $f_g \in \mathbb{R}^{1\times 6d}$. To increase the capacity of the model, more blocks can be stacked. Suppose $M$ blocks are used, the concatenated feature $f_c$ will have dimensions $\mathbb{R}^{N \times (M+1)d}$ and yield $f_g \in \mathbb{R}^{1\times 2(M+1)d}$.

\subsubsection{Regression Head}
The regression head adopts two-layer MLP similar to the embedding module. The first linear layer decreases dimension of the global feature $f_g$  to its half. After LayerNorm and SiLU activation, the second linear layer maps the half dimension to 1 scalar as the estimation of  $\hat{H}_s$. The expression is denoted as:
\begin{equation}
    \hat{H}_s = \textrm{Linear}\left(\textrm{SiLU}\left(\textrm{LayerNorm}\left(\textrm{Linear}.(f_g)\right)\right)\right)
\end{equation}

The network is trained using Huber loss\cite{huber1964}:
\begin{equation}
    L(x) = \begin{cases}
        \frac{1}{2}x^2, & |x| < 1, \\
        |x| - \frac{1}{2}, & \text{otherwise},
    \end{cases}
\end{equation}
where $x = H_s - \hat{H}_s$ represents the error between the true and estimated  heights.  Here, Huber Loss replaces the mean squared error (MSE) criterion by a linear penalty for those large outliers, while maintaining the smooth optimization benefits of a quadratic penalty for small errors.

\section{Experiments}
\subsection{Experimental Setup and Radar Parameters}

\begin{figure}[!ht]
    \centering
    \includegraphics[width=1.\linewidth]{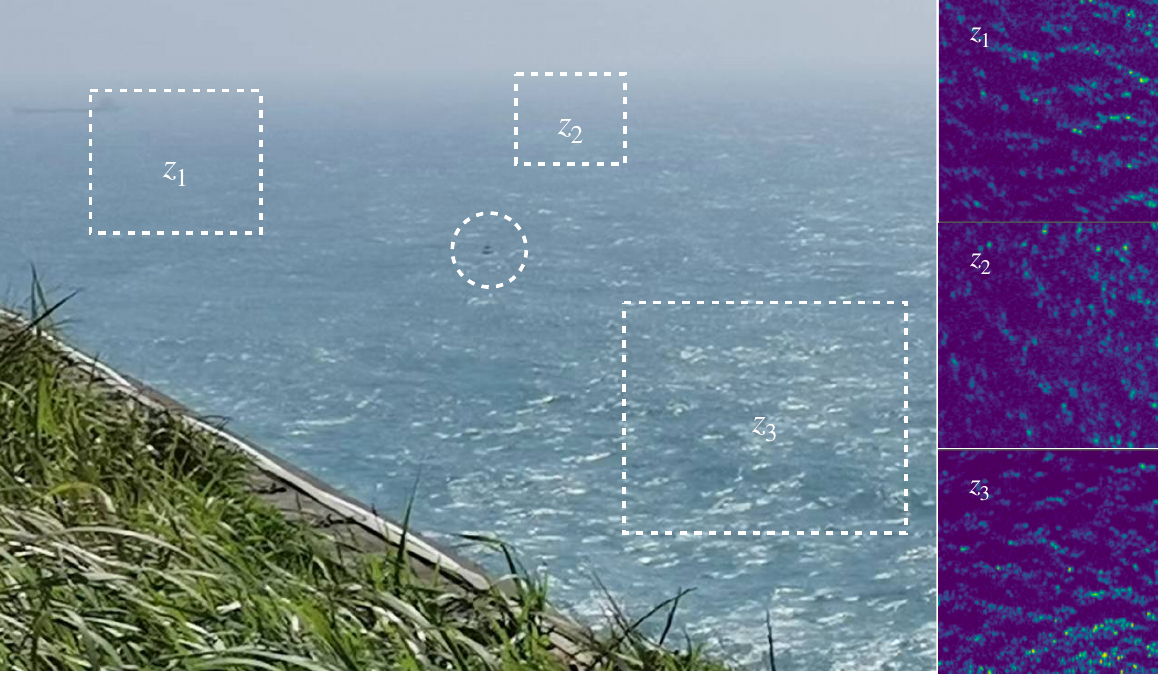}
    \caption{On-site experimental environment as viewed from the radar location. The white-dashed circle marks the wave-measuring buoy, and the rectangles indicate three observed sea zones ($z_1, z_2, z_3$), with their corresponding raw sea clutter images displayed on the right.}
    \label{fig.env}
\end{figure}

The significant wave height data used in this experiment were collected from a  wave-measuring buoy near Shengshan Manyutou Island, East China Sea (Fig. \ref{fig.env}). The $H_s$ updating interval was 30 minutes, with a measurement precision of $\pm (0.1+H_s\times 10\%)$ meters, and the local water depth was approximately 39 meters. The marine radar used for observation had a 2-meter antenna mounted on a hilltop about 50 meters above sea level, rotating at a speed of 2.5 seconds per revolution to collect echoes. The radar's beam width was 1 degree and its range resolution was 7.5 meters.

The marine radar captured planar coordinate images (128x128) in three directions around the buoy—north, east, and south—and saved them as image sequences limited to 64 frames per sequence. The observed range of these images spanned approximately 1.5 km to 2.5 km from the radar location. Data collection took place between July and October 2021, resulting in a data set consisting of 1620 image sequences. During this period, the experiment also encountered extreme weather conditions with wind speeds exceeding 20 m/s.

The sea clutter data set includes significant wave heights  ranging from 0.5  to 6.3 meters. 80\% of the heights are less than 2.5 m and 2\% of them are higher than 4 m. Since simultaneously three sea clutters are captured in zones 1, 2 and 3, the training data and testing data are randomly chosen from these three zones with ratios of 2:1 at each recording time (results in 1080 training samples and 540 testing samples). This sampling scheme makes sure that the heights distributions are equally in both the training and testing sets.

\subsection{Model Comparison}

This study compares SWHFormer \cite{swhformer2024}, a vision transformer (ViT)  \cite{vit2021} based architecture to regress a significant wave height from a single image of sea clutter. Since the  acquired images have a resolution of 128 $\times $128, while SWHFormer was fine-tuned from ViT pre-trained on 384 $\times $384 range-azimuth images, two versions are tested for fairness: SWHFormerI and SWHFormerII. SWHFormerI adopts the same parameter settings and pre-trained weights as  \cite{swhformer2024}, requiring input images to be upsampled to 384 $\times $384 via preprocessing. In contrast, SWHFormerII, while sharing the same architecture, is trained from scratch using 64 $\times $128 $\times $128 spectral amplitude data fed into the ViT model. Comparison with SWHFormerII, which utilizes all spectral points, allows evaluating whether the sparse frequency-point-based SPT can compete with general vision model using full spectral data. Additionally, this work compares 3D-VGG  \cite{3dvgg2023}, a 3D convolutional network for regressing SWH from image sequences. 3D-VGG employs classic VGG convolutional neural network modules \cite{vgg2015} to learn spatiotemporal kernels that capture the relationships between time-frequency information and wave heights, thus testing the effectiveness of using different architectures. For sparse spectral-point data, we compare Naive Point Transformer (NPT) \cite{attention2017} and SPCT (Simple Point Cloud Transformer)  \cite{pct2021} to validate the need to select the point cloud feature and adapt module. Here, NPT treats the power  $p $ of each spectral point as a token, region and coordinate vectors ( $[z,k\_x,k\_y, w] $) are independently embedded in additional positional embedding module and uses a class-token to learn representative features for $H_s$ regression. SPCT uses the same input feature as SPT,  but adopts the convolution embedding, bacth normalization, and offset attention modules.

Metrics to test the performance of different models  include the number of parameters, FLOPS (floating-point operations per second) for a sample, Pearson Correlation Coefficient (CC) between the estimation  and the heights of the ground truth, the root mean square error (RMSE) for  low ($H_s$ in the range 0.5-2.5 m), medium (2.5-4.0 m),  high sea states ( $>$4 m) and  overall $H_s$ estimation.

\begin{table*}[ht]

    \footnotesize 
    \caption{Performance comparison of different models}
    \begin{tabular}{@{}lc c c c  c c c c@{}}
        \toprule
        \multirow{2}{*}{Model} & \multirow{2}{*}{Input} & \multirow{2}{*}{Parameters}& \multirow{2}{*}{FLOPS}  & \multirow{2}{*}{CC} & \multicolumn{4}{c}{RMSE(m)$\downarrow$  }\\
        \cmidrule(lr){6-9}
        & Feature & (M)$\downarrow$ & (G)$\downarrow$  & $\uparrow$ & overall & 0.5-2.5 m & 2.5-4.0 m & $>4$ m\\
        \midrule
        SWHFormerI\cite{swhformer2024} & single image & 86.4 & 49.7 & 0.85 & 0.51 & 0.47 & 0.69 & 1.63 \\
        SWHFormerII & 3D spectrum & 47.9 & 3.40 & \textbf{0.97} & 0.24 & 0.22 & \textbf{0.31} & 0.49 \\
        3D-VGG\cite{3dvgg2023} & image sequence& \textbf{0.43} & 94.6 & \textbf{0.97} & 0.25 & 0.19 & 0.33 & 0.73 \\
        NPT\cite{attention2017} & spectral points& 0.70 & 1.04& 0.71 & 0.79 & 0.76 & 0.61 & 1.75 \\
        SPCT\cite{pct2021} & spectral points& 0.64 & 1.54 & 0.96 & 0.25 & 0.20 & 0.36 & 0.66 \\

        SPT (Ours) & spectral points& 0.70 & \textbf{1.04} &  \textbf{0.97} & \textbf{0.23} & \textbf{0.17} & 0.34 & \textbf{0.48}\\
        \bottomrule
    \end{tabular}
    \medskip
    \centering
     \\Lower values for Parameters, FLOPS and RMSE are better, while higher values for CC are better. 
     \\Bold numbers indicate the best performance in each column.
    \label{tab:comparison}
\end{table*}

In Table~\ref{tab:comparison}, SPT demonstrates superior performance in several critical terms, including computing FLOPS, CC, and overall RMSE. The results show that

\begin{itemize}
    \item SPT achieves the best balance between model complexity and performance, with minimal parameters (0.70 M) and FLOPS (1.04 G).
    \item  While 3D-VGG has fewer parameters than SPT, its computational cost is significantly higher due to 3D convolutions.
    \item SPT outperform SWHFormerII which uses all spectrum data, indicating that sparse spectral points are sufficient for effective significant wave height estimation.
    \item SPCT and SPT outperform NPT. It suggests that positional information should be viewed as the initial feature to enhance the discrimination of the energy distribution in the spectrum.
    \item  Compared to SPCT, SPT's MLP embedding and self-attention with QK Norm are more suitable to the structure of  point cloud in spectrum.
\end{itemize}

The experimental results validate the effectiveness of our proposed method in capturing essential features from sparse spectral data while maintaining computational efficiency.

\subsection{Analysis of SPT Characteristics}
This section discusses two key characteristics of the SPT model: training efficiency and the alignment of learned features with  dispersion relation.

(1) \textbf{Training Efficiency}:

Table~\ref{tab:timecost} presents comparison of per-batch computational performance for the 3D-VGG, SWHFormerII, and SPT based on image sequences, full spectra, and sparse spectral points, respectively. Evaluations are conducted on an RTX 5090 GPU with 32 GB graphic memory.     
       \begin{table*}
        \centering
        \footnotesize  
         \caption{GPU training profile for a batch}
        \begin{tabular}{ccccclcc}
    \toprule
 Model& Load to & Load to  & Forward & Backward&  Time cost &GPU  & Memory\\
      &   Memory(ms) $\downarrow$ & GPU(ms)$\downarrow$ & (ms)$\downarrow$ & (ms)$\downarrow$& (ms)$\downarrow$ & Utilization $\uparrow$  & Utilization$\downarrow$  \\
      \midrule
         3D-VGG&  195&  18.2&  138&  488&   839& \textbf{74\%} & 98\% \\
            \midrule
         SWHFormerII&  198&  19.5&  5.81&  23.7&   247& 6\%& 8\%\\
            \midrule
         SPT&  \textbf{0}&  \textbf{5.01}&  \textbf{4.18} &  \textbf{9.62}&   \textbf{18.8}& 70\% &  \textbf{3\%}\\
         \bottomrule
    \end{tabular}
    \label{tab:timecost}
    \end{table*}

Due to the large size of the 1620 sequence dataset (approximately 26 GB), sequential images or full 3D spectrum must be loaded into host memory and GPU memory in batches. Among the three methods, 3D-VGG requires the highest computational load, and its batch size is limited by the GPU memory capacity. In this study, a batch size of 32 is adopted for 3D-VGG, at which the GPU memory utilization is nearly saturated. Other two methods set the batch size as 32 for fair comparison.   

Both 3D-VGG and SWHFormerII need to load sea clutter data from the hard disk in batches, requiring over 190 ms to transfer data from disk to host memory. Although the computational cost of 3D convolution is approximately four times higher than that of the lightweight vision transformer used in SWHFormerII, its overall regression accuracy does not surpass that of SWHFormerII.

SPT employs sparse frequency-point data, which significantly reduces the data loading time from disk. During preprocessing, SPT selects only the strongest 0.05\% of spectral points for significant wave height estimation, which is key to its computational efficiency. Since all sparse data can be loaded into GPU memory at once, the communication overhead among hard disk, host memory, and GPU memory during batch processing is minimized, thereby improving GPU utilization and saving considerable training time. For SPT, the main data movement occurs between GPU memory cache and CUDA cores, taking only about 5 ms.

SPT exhibits a significantly lower per-batch training time. It is approximately 44$\times$ and 13$\times$ faster than 3D-VGG and SWHFormerII, respectively. In terms of GPU utilization efficiency, 3D-VGG maintains high GPU activity because computation proceeds concurrently with data loading; however, its memory bandwidth usage reaches 98\%, nearing the limit. For SWHFormerII, the data loading time far exceeds the time for forward and backward propagation, resulting in considerable GPU idle periods. Its 6\% GPU utilization suggests potential for acceleration through improved data preprocessing and caching strategies. In SPT, data transfer within GPU memory is comparable to the forward propagation time, yielding an overall GPU computational efficiency of 70\%.

With the total number of epochs set to 300 and 34 batches of training data, SPT can complete training within 4 minutes. Its memory bandwidth usage is only 3\%, indicating that the SPT algorithm could scale to datasets ten times larger while still being trainable on a single GPU. The results show that SPT achieves superior computational efficiency with minimal memory overhead, making it suitable for real-time applications.

\begin{figure}[!ht]
            \centering
            \includegraphics[width=1.\linewidth]{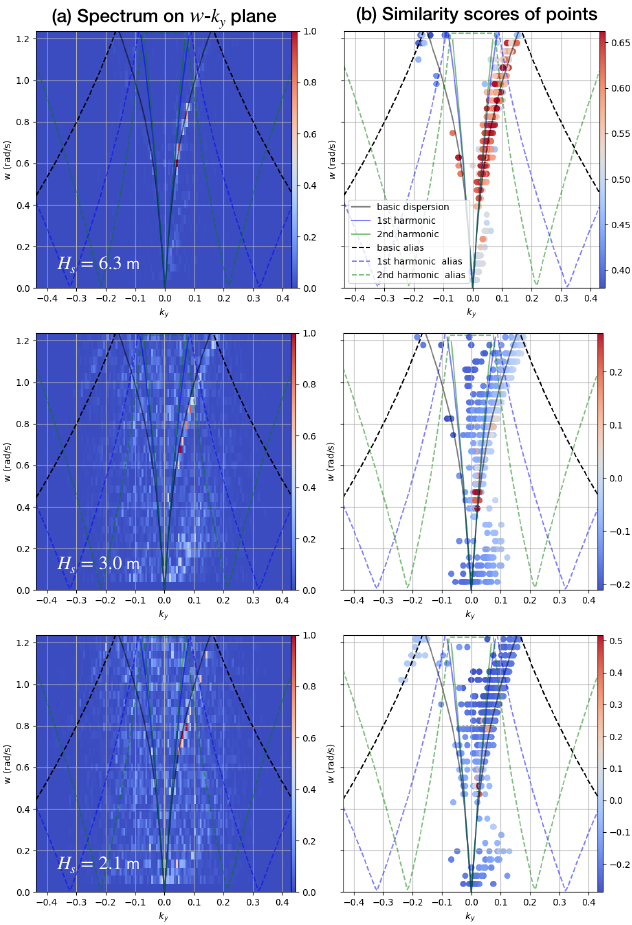}
            \caption{Alignment of SPT features with physical dispersion relations. The left column shows $w$-$k_y$ planes for different $H_s$ values. Solid lines represent dispersion relations, and dashed lines indicate aliasing effects. The right column highlights feature similarity scores, with higher intensities indicating stronger influence on regression.}
            \label{fig.align}
        \end{figure}

(2) \textbf{Physical Dispersion Alignment}:
        
        Figure~\ref{fig.align} illustrates the alignment of SPT-learned features with physical dispersion relations. The left column displays the $w$-$k_y$ plane for wavenumber-frequency spectra at different significant wave heights ($H_s$). Solid lines represent fundamental and harmonic dispersion relations, while dashed lines indicate frequency aliasing effects. Right column shows the Cosine similarity between the global feature $f_a$ and the local feature of each selected spectral point. Higher similarity score means more impact to regression.

In figure \ref{fig.align}, high sea states ($H_s$=6.3 m) exhibit energy concentration along the dispersion curves, which is mirrored in SPT's learned features. The points with strong similarities are densely clustered along the dispersion curves also. As the significant wave height decreases and the wave period shortens, spectral interference becomes more pronounced. The spectral points that play a decisive role in the final discrimination become sparser, yet their positions of high similarity remain consistent with the dispersion curve. Taking the last sub‑figure with a significant wave height of 2.1 m as an example, the points with high feature‑similarity scores are located at angular frequencies $w = 0.8$ and 0.5 rad/s, corresponding to wave periods of 7.9 s and 12.6 s, respectively. The 7.9 s period closely matches the buoy‑observed value of 7.8 s, while the longer 12.6 s period indicates that SPT learns to utilize longer‑period wave textures to further distinguish subtle differences among low sea‑state data.

\subsection{Ablation Study}\label{subsec.abl}
To identify optimal configurations, we conducted ablation studies across four key aspects: input features, attention mechanisms, query-key normalization, and model scalability.
Numerical results are listed in Table \ref{tab:ablation}.
\begin{table*}[!ht]
    \centering
    \caption{Ablation study results for SPT hyper-parameters.} \label{tab:ablation}
    \footnotesize
    \begin{tabular}{c c c c c c c c}
       \toprule
        \multirow{2}{*}{Hyper} & \multirow{2}{*}{Setting} & \multirow{2}{*}{CC} & 
        \multicolumn{4}{c}{RMSE (m) $\downarrow$} \\
        \cmidrule(lr){4-7}
        parameters& options & $\uparrow$ &  overall & 0.5-2.5 m & 2.5-4.0 m & $>4$ m\\
        \midrule
        input features &[$w, p$]                   & 0.93 & 0.34 & 0.26 &0.45 & 1.14 \\
        input features &[$k_x, k_y, w$]          & 0.96 & 0.24 & 0.18 &0.39 & 0.80 \\
        input features &[$z,k_x, k_y, w, p$]  & \textbf{0.97} & 0.23 & 0.17 &0.34 & 0.48 \\
        
        \midrule
        attention type& offset   & 0.96 & 0.24 & 0.17 & 0.36 & 0.78  \\
        attention type& residual & \textbf{0.97} & 0.23 & 0.17 & 0.34 & 0.48  \\
        \midrule
        QK Norm.& No & 0.96  & 0.24 & 0.18 & 0.34 & 0.70  \\
        QK Norm.& Yes & \textbf{0.97} & 0.23 & 0.17 & 0.34 & 0.48  \\
        \midrule
        $[N, \text{heads}, d, \text{blocks}]$ & $[256,4,64,1]$      & 0.95  & 0.27 & 0.19  & 0.36 & 1.04\\
        $[N, \text{heads}, d, \text{blocks}]$ & $[512,8,128,2]$     & 0.97 & 0.23 & 0.17  & 0.34 & 0.48\\ 
        $[N, \text{heads}, d, \text{blocks}]$ & $[1024,16,256,4]$ & \textbf{0.97} & \textbf{0.22} & \textbf{0.17}  & \textbf{0.31} & \textbf{0.43} \\
        \bottomrule
    \end{tabular}
\end{table*}

Key findings include:
\begin{itemize}
    \item \textbf{Input Features}: Incorporating zone identifiers ($z$)  and coordinates ($k_x, k_y, w$) enhances model performance by leveraging spatial context and wavenumber coordinates. Based on empirical experience in retrieving wave height from sea clutter, the location of the maximum power spectral point $w_p$ and the distribution pattern of other high‑power points in wave-number spaces are related to the wave energy. Moreover, radar echoes captured from different directions exhibit different power levels, leading to energy differences in sea clutter from distinct regions at the same time (a phenomenon also discussed in \cite{Lund2014}). Incorporating these two pieces of prior knowledge into the feature representation can help the model achieve better training results.
    \item \textbf{Attention Mechanisms}: Residual attention outperforms offset attention used in \cite{pct2021} due to better handling of sparse spectral data and improved numerical stability. There are two main reasons for this:  
(1) The offset attention in PCT employs 1D convolutions for dimension expansion. While convolutions can effectively extract discriminative features from local neighborhoods in dense point cloud with continuous geometric distributions (e.g., point cloud of airplanes or cars), they tend to suppress the characteristics of independent strong-valued points in sparse spectral point data.  
(2) PCT uses batch normalization (BatchNorm) for numerical stability. Due to the limited number of high sea-state samples  collected during typhoon periods in this dataset, high sea-state data are significantly outnumbered by medium and low sea-state samples within each training batch. The averaging operation in BatchNorm severely weakens the discriminative ability of high sea-state features for wave height estimation. Consequently, SPT which adopts MLP-based dimension expansion with standard self-attention and LayerNorm improves the overall performance of the model.
    \item \textbf{QK Normalization}: Applying root mean square normalization stabilizes training, particularly for high sea states with limited sample sizes. Considering that the power $p$ of points can vary by up to two orders of magnitude within the same spectrum and more across different sea states, introducing root-mean-square normalization before computing the inner product of the query and key in the attention mechanism helps to stabilize the numerical range, prevents drastic fluctuations in back-propagated gradients, and thus enables better convergence of the model.
    \item \textbf{Model Scaling}: Increasing model capacity (larger $N$, heads, embedding dimension $d$ and transformer blocks) improves performance, consistent with transformer's scaling laws~\cite{scalinglaw2020}.
\end{itemize}

These insights guide the selection of optimal configurations for effective wave height estimation.

\section{conclusion}

This work provides an efficient framework for the estimation of $H_s$ from sparse spectral data and offers insights into the application of transformer models in learning spectral discrimination. The proposed method not only aligns well with the physical dispersion relations but also demonstrates practical utility for real-time wave monitoring systems.  However, there remains a limitation. The current dataset is limited in size and diversity, particularly from the site of  a nearshore island. 
Future research directions include collecting more diverse training data to further explore its applicability in different environmental conditions and extending the model to other oceanographic parameters, such as current speed and direction.

\section{Acknowledgment}
The authors thank Zhiwei He (SVA Communication Technology Co., Ltd.) for preparing the sea clutter datasets and for technical discussions during the experiments.
\bibliographystyle{IEEEtran}
\bibliography{tgars2026_ref_2026} 

\begin{IEEEbiography}{Yi Zhou}
received the B.S. and M.S. degrees in Electronic Engineering from Dalian Maritime University, China, in 2003 and 2006, respectively. In 2012, he obtained a joint Ph.D. degree in Signal Processing from Shanghai Jiao Tong University, China, and the University of Technology of Troyes, France. Since 2013, he has been a lecturer in the Department of Electronic Information Engineering at Dalian Maritime University. During the winter of 2021, he was a visiting researcher with the TSAIL group at Tsinghua University for three months. His research interests include marine radar signal processing and computer vision.
\end{IEEEbiography}

\begin{IEEEbiography}{Li Wang}
was born in Yibin, Sichuan, China, in 1984. He received the Ph.D. degree from Tsinghua University, China, in 2019. He is currently a Senior Engineer at SVA Communication Technology Co., Ltd. His research interests mainly include adaptive radar signal processing and radar waveform design.   
\end{IEEEbiography}

\begin{IEEEbiography}{Hang Su}
(IEEE member), is an associate professor in the Department of Computer Science and Technology at Tsinghua University, Beijing, China. He received his Ph.D. degree from Shanghai Jiao Tong University in 2014 and worked as a visiting scholar at Carnegie Mellon University from 2011 to 2013. 
His research interests include the adversarial machine learning and robust computer vision. He has served as area chair for NeurIPS and as a workshop cochair in AAAI 2022. He received ``Young Investigator Award” from MICCAI2012, the ``Best Paper Award” in AVSS2012, and ``Platinum Best Paper Award” in ICME2018.
\end{IEEEbiography}

\begin{IEEEbiography}{Tian Wang}
received the M.S. degree from Xi’an Jiaotong University, China, in
2010, and the Ph.D. degree from the University of Technology of Troyes, France, in 2014. He is a professor at the School of Artificial Intelligence, Beihang
University. His research interests include computer vision and pattern recognition.
\end{IEEEbiography}

\end{document}